\documentstyle[titlepage,12pt]{article}
\begin{document}
\parindent 2em

\begin{titlepage}
\begin{center}
\vspace{12mm}
{\LARGE Finite temperature transport at the
superconductor-insulator transition in disordered systems}
\vspace{15mm}

Igor F. Herbut

Department of Physics and Astronomy, University of British Columbia,\\
6224 Agricultural Road, Vancouver B. C., Canada V6T 1Z1, and\\
$^*$ Department of Physics, Dalhousie University, Halifax, Nova Scotia,
Canada B3H 3J5\\
\end{center}
\vspace{10mm}

{\bf Abstract:} I argue that the incoherent, zero-frequency limit of the
universal crossover function in the temperature-dependent
conductivity at the superconductor-insulator transition in disordered
systems may be understood as an analytic function of dimensionality
of the system $d$, with a simple pole at $d=1$. Combining the exact
result for the crossover function in $d=1$ with the recursion relations 
in $d=1+\epsilon$, the leading term
in the Laurent series in the small parameter
$\epsilon$ for this quantity is computed for the systems of disordered
bosons with short-range and with Coulomb interactions. The universal,
low temperature, dc critical conductivity for the dirty boson 
system with Coulomb interactions in $d=2$ is estimated 
to be $\sim 0.69 (2e)^2 /h$, in good agreement with many experiments on
thin films. The next order correction is likely to somewhat increase the
result, possibly bringing it closer to the self-dual value.
\vspace{100mm}
\end{titlepage}


  Different quasi two-dimensional (2D) electronic systems, like
thin films \cite{liu}, Josephson junction arrays \cite{zant} or  underdoped
high-Tc cuprates \cite{fukuzumi}, appear to have a continuous 
zero-temperature phase transition
between the superconducting and the insulating state, 
as some parameter of the system is varied.
The loss of phase coherence in the  ground state
is believed to be due to Anderson localization of the 
Cooper pairs \cite{ma}, \cite{fisher}, which through the quantum uncertainty
relation competes with the  phase ordering.   In reality, 
this purely quantum ($T=0$) phenomenon is unavoidably observed at low but 
finite temperatures, typically as a  crossover
from increasing to decreasing dc conductivity  with 
temperature. Particularly interesting is the
behavior of the conductivity near the critical value of the tuning
parameter, which remains finite down to the lowest temperatures,
and usually very close to $(2e)^2 /h$, the quantum unit of conductance for 
electron pairs. This near-universality of
the metallic transport right at
the superconductor-insulator (SI) transitions in 2D has been
a subject of numerous theoretical and experimental investigations 
within the last decade. Scaling 
arguments \cite{girvin} imply that at $T=0$ and right at the quantum
critical point the dc conductivity must indeed be universal,
very much like the critical exponents.   
This insight has received a substantial amount
of theoretical support in form of concrete numerical \cite{cha},
\cite{batrouni}, \cite{wallin} and analytical \cite{cha}, \cite{fazio}, 
\cite{herbut} evaluations of the $T=0$ critical conductivity
for various universality classes of the SI transitions.
The experimental evidence for the expected universality of the
charge transport however, is still 
somewhat less convincing. In particular, measurements  typically 
yield values of the critical conductivity two to three times larger 
than the calculations \cite{liu}. Most of the theoretical studies
\cite{cha}, \cite{batrouni}, \cite{wallin}, \cite{fazio}, \cite{herbut}
of the critical conductivity however, are in  significant
conceptual discord with the experiment, as recently emphasized by  
Damle and Sachdev \cite{damle}. They argued  
that the $\omega=0$, $T\rightarrow 0$  dc conductivity that is typically
measured and the $T=0$, $\omega\rightarrow 0$ conductivity that is
usually calculated, originate in entirely different
dissipation mechanisms, and while both should be universal, they 
have no reason to be equal. As an illustration, they 
obtained the conductivity in the hydrodynamic, incoherent transport regime 
$\hbar \omega/k_B T \rightarrow 0$ at the simpler
superfluid-Mott insulator (SF-MI) transition brought by an external
periodic potential, near its upper critical
dimension of $d=3$, and
indeed found a different, and larger, value of the critical conductivity.

 In real disorder systems the insulating phase is presumably
the result of Anderson localization  and should therefore be
a compressible Bose-glass (BG) \cite{fisher},
not the Mott insulator. It has been argued
that, with disorder present, the limits $\omega\rightarrow 0$ and
$T\rightarrow 0$ may commute \cite{cha}, and the imaginary time
Monte Carlo calculations
indeed found little or no dependence of the critical conductivity at the
SF-BG transition on the ratio $\omega_n /T$ \cite{wallin},
where $\omega_n$ is the Matsubara frequency. Including
disorder makes the problem difficult to study analytically, 
since then the transition seems to lack the upper critical dimension,
or any other obvious limit in which 
at the criticality the system would be weakly coupled. Numerical
techniques \cite{cha}, \cite{batrouni}, \cite{wallin}
on the other hand, determine the
conductivity by extrapolating from the
imaginary Matsubara frequencies, $\omega_n = n 2\pi k_B T$,
and thus by its nature are 
not very sensitive to any structure it might have in the
hydrodynamic regime. The understanding of the role of
disorder and the calculation of the
experimentally measured low temperature  dc  
conductivity at the SI critical point in 2D systems thus 
presents itself as a fundamental and unsolved problem.
The purpose of this Letter is to propose its solution 
in form of a controlled expansion around the solvable case
of the SF-BG transition in one dimension (1D). 
 
I will consider the system of {\it disordered} interacting bosons
of charge $e_* = 2e$ defined, for example, by the Bose-Hubbard,
or Josephson junction array Hamiltonian
with a random chemical potential \cite{fisher}, \cite{herbut}.
This model is known to possess a continuous SF-BG transition at
$T=0$, and should be appropriate for description of
the SI transition observed in Josephson junction arrays,
or $^4 He$ in random media.
If the Coulomb interaction between bosons is added, it should
represent the correct universality class for the 
transition in homogeneous and granular films.
Although there is no upper critical
dimension for the SF-BG transition in the usual sense 
\cite{weichman}, \cite{fisher}, \cite{herbut1}
the fact that the superfluid phase
in $d=1$ and at $T=0$ exhibits only a power-law long-range order
implies that $d=1$ represents the lower
critical dimension. Recently, this observation
has been used by the author to formulate 
a controlled expansion of the universal quantities at the SF-BG
transition in powers of a small parameter $\epsilon=d-1$
\cite{herbut1}. In particular, and as argued below,
in the limit $T\rightarrow 0$
the dc resistivity at the critical point  
is proportional to a certain power of temperature and to the value of
the disorder parameter at the SF-BG critical point, which becomes
infinitesimally small
($\sim \epsilon$) as dimensionality of the system is reduced to 
$d=1$. This suggests that the universal part of the low temperature dc
conductivity at the critical point may be expressed
as an analytic function of
$\epsilon$, with a simple pole at $\epsilon=0$. I compute
the first term in the Laurent series for the real part
of the critical dc conductivity around $\epsilon=0$:
\begin{equation}
\sigma'_c(\omega=0,T)
=(\frac{\hbar c}{k_B T})^{\frac{2-d}{z}} [ \frac{6 x}{\pi^{5/2}}
\frac{1}{\epsilon} +O(1) ]  \frac{e_* ^2}{h},
\label{1}
\end{equation}
where $c$ is a microscopic constant with units $(length)^z /time$,
$z=\{d,1 \}$ is the dynamical exponent
and $x=\{1,2 \}$ for the short-range and the Coulomb
interaction between bosons, respectively. For $d=2$ this leads to an 
estimate $\sigma_c \approx 0.69 e_{*} ^2 /h$
for the Coulomb universality class,
in agreement with many experiments on thin films \cite{liu}. 
A systematic perturbative procedure for a further calculation of the
critical conductivity and of the critical exponents is outlined,
and the sign of the next  order
correction in the Eq. (1) is discussed. I speculate 
that at the SF-BG critical point in 2D the real part of the conductivity is a
continuously decreasing function of the ratio $\hbar \omega/k_B T$,
qualitatively similar to the result in $d=1+\epsilon$. 
 
   In general, right at the critical point in a
 $d$-dimensional system the real part of conductivity for the 
 frequencies and the temperatures much smaller then some microscopic
 cutoff energy can be written as
 \begin{equation}
 \sigma'_c(\omega,T)= (\frac{\hbar c}{k_B T})^{\frac{2-d}{z}}  F_d
 (\frac{\hbar\omega}{k_B T}) \frac{e_* ^2}{h},
 \label{2}
 \end{equation}
 where $F_d (x)$ is a {\it universal} crossover function. While I focus  
 here on the SI transition, the above equation
 applies equally well at the
 critical points in quantum Hall systems \cite{sondhi} and
 doped semiconductors \cite{lee}, for example. 
 At $d=2$ the power of the temperature-dependent
 dephasing length \cite{sondhi} in the last equation vanishes,
 and the conductivity in units of $e_* ^2 /h$
 becomes a universal function of the
 ratio of quantum and thermal energy scales.
 It then becomes evident that, in principle, 
 at the critical point the conductivity
 in $d=2$ assumes two completely different values depending on the order of
 limits $\omega\rightarrow 0$ and $T\rightarrow 0$, that correspond
 to the values of the crossover function either at zero or at infinity, and
 measure completely incoherent or completely
 coherent transport, respectively. What is more
 surprising, although it arises as a natural consequence
 of the scaling law, is that even in the limit $\hbar \omega
  / k_B T \rightarrow 0$ of incoherent, collision dominated transport,   
 conductivity at the critical point in $d=2$ is still given by
 a {\it universal} number
 \cite{damle}. To understand the dependence of the conductivity in this
 limit on dimensionality of the system, consider 
 the effective low-energy action for the disordered
 superfluid that describes the SF-BG transition
 in the dirty Bose-Hubbard model in $d=1$ 
 \cite{giamarchi}, \cite{herbut}, \cite{herbut1}:
 \begin{eqnarray}
 S=\frac{K}{\pi} \sum_{i=1}^{N} \int dx \int_{0}^{\beta} d\tau
 [ c^2 (\partial_x \theta_i(x,\tau))^2 +
 (\partial_\tau \theta_i(x,\tau))^2 ] \nonumber \\  
 -D \sum_{i,j=1}^N \int dx \int_{0}^{\beta} d\tau d\tau'
 \cos 2(\theta_i (x,\tau) - \theta_j (x,\tau')), 
 \label{3}
 \end{eqnarray}
 where $\beta=\hbar/k_B T$, $K$ is inversely proportional to the superfluid
 density, $c$ is the velocity of low-energy phononic excitations,
 index $i$ numerates the replicas introduced to average over disorder, and
 $D$ is related to the width of the distribution of the random potential. 
 The usual limit $N\rightarrow 0$ and
 a short-distance cutoff $\Lambda^{-1}$ in the Eq. (3) are
 assumed. This effective action arises as
 the one dimensional realization of the density (dual) representation 
 of the dirty Bose-Hubbard model at low energies \cite{herbut}, or
 as the bosonic representation of the
 disordered Luttinger liquid \cite{giamarchi}.
 Invariance under a change of the cutoff implies
 that the conductivity in $d=1$ can be written in the scaling form
 \begin{equation}
 \sigma(\omega,T)=\frac{\hbar c}{k_B T} f(K(b),
 c(b), D(b),
 \frac{\hbar \omega}{k_B T} ) \frac{e_* ^2}{h},
 \label{4}
 \end{equation}
 where $K(b)$, $c(b)$ and $D(b)$ are the renormalized couplings at the new
 cutoff $b \Lambda^{-1}$, and $b=\hbar c\Lambda/k_B T$.
 At low temperatures the conductivity is determined by the infrared stable
 fixed point of the scaling transformation. The result of the renormalization
 in the theory (3) for weak
 disorder is well known \cite{giamarchi}, \cite{herbut1}:
 under the change of cutoff the combination $\kappa=Kc^2$, that is
 proportional to the compressibility, stays constant,
 $K$ always increases,
 and small $D$ is relevant for $\eta=Kc>1/3$, and irrelevant
 otherwise. There exists a separatrix in the $\eta-D$
 plane which ends in the SF-BG critical point at
 $\eta=1/3$ and $D=0$. At $\omega=0$  
 the scaling function in the Eq. (4) at weak disorder should behave as
 $f\sim 1/D(b)$ \cite{giamarchi},  \cite{herbut2}, 
 and therefore right at the separatrix slowly (logarithmically, 
 $\sim (\ln(b))^2 $,  \cite{herbut2})
 diverges as $b\rightarrow \infty $ in $d=1$. Apart from the
 logarithmic correction that derives from disorder being dangerously
 irrelevant at the SF-BG criticality in 1D, right at the transition
 the Eq. (4) is just a special case of the general scaling form (2), for
 $d=z=1$. Since in $d=1+\epsilon$
 disorder at the transition scales towards a finite fixed point value
 $D(b) \rightarrow D^* \sim \epsilon$,
 when $b\rightarrow \infty $ \cite{herbut1}, the comparison of the two 
 scaling laws in Eqs. (2) and (4) leads to the identification
 \begin{equation}
 F_{1+\epsilon}(0) = f(1/3,1, D^*, 0) = \frac{const.}{\epsilon} + O(1),
 \label{5}
 \end{equation}
 which expresses the central idea of this work. 
 The leading term in the Laurent series for $F_d (0)$
 is completely determined
 by the scaling function as in 1D and by the infinitesimal
 value of disorder at the
 fixed point of the scaling transformation in $d=1+\epsilon$.
 
 In the remaining of the paper the complete function $f$
 near the SF-BG critical point in $d=1$ is obtained,
 a new field-theoretic version of
 the recursion relations in $d=1+\epsilon$ requisite for determination
 of the fixed point value of disorder is derived, and finally,  
 the residuum at the pole at $\epsilon=0$ in the Eq. (5) for both 
 short-range and Coulomb interactions between bosons is computed. 

 The standard linear response formalism
 yields the conductivity in $d=1$
 \begin{equation}
 \sigma(\omega,T)=-i \frac{2\omega}{\pi} \frac{e_* ^2}{h}
 \lim_{N\rightarrow 0} \frac{1}{N} \sum_{n,m=1}^N G^r _{nm}(\omega). 
 \label{6}
 \end{equation}
 $G^r_{nm}(\omega)$ is the
 temperature dependent, retarded, 
 $q=0$ Green's function defined as
 \begin{equation}
 G^r _{nm}(\omega) = \int dx \int_0 ^{\beta} d\tau e^{i\omega_n \tau}
 \langle T_{\tau} \theta_n (x,\tau) \theta_m (0,0) \rangle_{i\omega_n
 \rightarrow \omega+i\delta},
 \label{7}
 \end{equation}
 where $T_{\tau}$ is the standard time-ordering operator. The thermal Green's
 function in the Eq. (7) may be evaluated perturbatively in disorder 
 and then analytically continued to real frequencies.
 Similar calculation has been performed before by Luther and Peschel
 \cite{luther} for $\eta>1$, which corresponds to weak coupling in the
 equivalent 1D fermionic system. The SF-BG transition at weak disorder is
 at $\eta\approx 1/3$, so I derive here a slightly improved version
 of their results which can be analytically continued 
 into the transition region. Introduce the self-energy as
 $G^t _{ij}(\omega_n) =\delta_{ij} / (2(K/\pi) \omega_n ^2 + \Sigma^t (
 \omega_n) )$. To the lowest order in $D$ it may be
 written as
 \begin{equation}
 \Sigma^t (\omega_n) = 8 D \int_0 ^{\beta} d\tau (1-e^{i\omega_n \tau})
 \langle T_{\tau} e^{i 2\theta_j (x,\tau)}
 e^{-i 2 \theta_j (x,0)} \rangle_0 + O(D^2),
 \label{8}
 \end{equation}
 where the average is performed over the quadratic part of the
 action (3). The self-energy is thus {\it itself} a Green's function, 
 which enables one to perform the analytic
 continuation to real frequencies by first rotating the
 integrand in (8) to real time by $\tau\rightarrow it$ to find the
 real time time-ordered propagator, and from it finally to determine
 the retarded one by using the standard relation between them \cite{mahan}.
 Performing the Fourier transform at the resulting expression then
 gives the retarded self-energy 
 \begin{equation}
 \Sigma^r (\omega) = \frac{16 \pi D}{c\Lambda}
(\frac{\pi k_B T}{\hbar c\Lambda})^{\frac{1}{\eta}-1} \sin(\frac{\pi}{2\eta})
e^{\frac{C}{\eta}} \int_{0}^{\infty} dt \frac{1-e^{i (\frac{\hbar \omega}
{k_B T})t}}{(\sinh(\pi t))^{\frac{1}{\eta}} }, 
\label{9}
\end{equation}
where $C\approx 0.577$ is the Euler's constant, and I assumed
that $\hbar c\Lambda /k_B T >> 1$, i.e. the
continuum (low-temperature) limit. Appearance of a  particular numerical 
constant is a consequence of the assumption that
the dispersion is $\omega= c k$ for all momenta $0<k<\Lambda$, and is
a non-universal, short-distance feature. This nevertheless,
does not compromise the universality of the conductivity at the
transition, since any non-universal constant like $C$ may at the end be 
absorbed into the definition of the running, dimensionless disorder 
coupling, as will be done shortly. The remaining 
integral in (9) is convergent only for $\eta>1$, but once  evaluated
there exactly, may be defined via
analytic continuation in the transition region $\eta \approx 1/3$.
Performing the integral, in the vicinity of
$\eta =1/3$ one  obtains the conductivity in $d=1$ to be 
\begin{equation}
\sigma(\omega,T) = \frac{i\omega c}{\eta(T)\omega^2 +
i 2 (k_B T/\hbar)^2 W(T)
g(\hbar\omega/k_B T) } \frac{e_* ^2}{h},
\label{10}
\end{equation}
where $W(T)=(\pi^4 D/c^2 \Lambda^3 e^{3C})(k_B T/\hbar c\Lambda)^{\frac{1}
{\eta }-3}$ is the dimensionless disorder variable,
$\eta(T)=\eta + W(T) \pi^{-3/2} \tan(\pi/2\eta)$, and
$g(x)=(1+(x/\pi)^2 ) \tanh(x/2)$. Note that the result indeed
may be cast into the scaling form as claimed in the Eq. (4).
The self-energy acquired an imaginary part, which in the dc limit
$\hbar\omega/k_B T\rightarrow 0$ becomes proportional to temperature
and to the temperature dependent disorder variable $W(T)$. The 
significance of the point $\eta=1/3$ now becomes apparent: for $\eta <1/3$
disorder variable $W(T)$ scales towards zero  
with decreasing temperature, and the
lowest order result in the Eq. (10) becomes asymptotically 
exact. At a finite
frequency and at $T=0$ the real part of conductivity in $d=1$ then becomes 
$\sigma'(\omega) = (\pi c/\eta(0))\delta(\omega)$ and the system
is an ideal conductor. If $\eta>1/3$ the
perturbation theory breaks down, which indicates the entrance into the
insulating phase. Notice that as $\eta \rightarrow 1/3^{-}$ the
coefficient in front of $W(T)$ in the expansion for $\eta(T)$ becomes 
divergent, as it has a simple pole at $\eta =1/3$. This is reminiscent 
of the dimensional regularization frequently employed in the studies of 
thermal critical phenomena \cite{binney}, and indicates that the theory
defined by the action in Eq. (3) becomes just renormalizable at $\eta =1/3$. 

 In the continuum limit, the effect of change of temperature on the 
 low-frequency conductivity in $d=1$ may be expressed entirely through the
 effective values of the coupling constants $\eta(T)$,
 $W(T)$, and, if we had retained the momentum dependence of the propagator,
 the compressibility $\kappa(T)$. This, of course, is just the statement of
 renormalizability of the theory, with the temperature used as an   
 infrared regulator. Taking into account the non-zero 
 canonical dimensions of the coupling constants away from $d=1$
  \cite{herbut1}, the effective couplings 
 close to $\eta=1/3$ satisfy the differential equations:
 \begin{equation}
 \dot{\eta}(T) = z^{-1} (d-1)\eta(T) -\frac{2}{\pi^{5/2}} W(T) + O(W^2 (T)), 
 \label{11}
 \end{equation}
 \begin{equation}
 \dot{W}(T) = (\frac{1}{\eta} -3) W(T) + O(W^2 (T)),
 \label{12}
 \end{equation}
\begin{equation}
\dot{\kappa}(T)=z^{-1} (d-z)\kappa(T), 
\label{13}
\end{equation}
 where $\dot{x}=dx / d\ln(k_B T/\hbar c\Lambda)$. The d-dependent
 terms in Eqs. (11) and (13) may be inferred 
 from the scaling of the superfluid density
 and the compressibility near the critical point \cite{fisher}, as
 discussed at length elsewhere \cite{herbut1}. Note that in
 the Eq. (13), unlike in the Eq. (11), there are no terms proportional to
 $W(T)$. This is a consequence of the exact symmetry of the
 interaction term in the action (2) under
 $\theta_i (x,\tau)\rightarrow \theta_i (x,\tau)+ h(x)$, for arbitrary
 function $h(x)$, and fixes
 the value of dynamical exponent to $z=d$ for the system
 with short-range interactions \cite{herbut1}. Linearization
 of the flow close to the fixed point of Eqs. (11) and (12) gives the
 correlation length exponent $\nu=(1/\sqrt{3\epsilon}) +O(1)$, in
 agreement with the result of the momentum-shell renormalization group
 \cite{herbut1}. The fixed point is located at $W^* (T)=\pi^{5/2}\epsilon /6 
 + O(\epsilon^2)$ and $\eta^* (T) = 1/3 +O(\epsilon)$. Using the Eqs.
 (10) and (5) one then obtains the main result announced in the
 Eq. (1), for the short-range interactions between bosons. 

   To make a comparison with experiments on thin films \cite{liu} or
on high-Tc cuprates \cite{fukuzumi} one needs
to include the long-range Coulomb repulsion between the electron 
pairs. A way to do this was proposed previously
by the author in the ref. 14, where the long-range interactions
was defined as $V(\vec{r})=e^2 \int d^d \vec{q}
\exp(i\vec{q}\cdot\vec{r}) / q^{d-1}$, so that it coincides with the
Coulomb interaction for $d>1$, and with the short-range interaction
precisely at $d=1$. The calculation of conductivity
in $d=1$ then remains the same,
the only change now being the equation for the 
temperature dependent charge $e^2 (T)$ instead for
the compressibility in the Eq. (13), with  $(z-d)\rightarrow (z-1)$,
and consequently, $(d-1) \rightarrow (d-1)/2$ in the Eq. (11) \cite{herbut1}. 
It then follows that with Coulomb interactions present
$z=1$, $\nu=\sqrt{2/3\epsilon}+O(1)$, and the fixed point value
$W^* (T) = \pi^{5/2}\epsilon / 12 + O(\epsilon^2)$.
This yields the second result in the Eq. (1), for the
Coulomb universality class. To the lowest order,
the critical dc conductivity is
larger if the Coulomb interaction is present. This may have been intuitively 
expected: a longer-range interaction suppresses the phase order
more efficiently, so it takes less disorder to finally  turn the system
into an insulator. 

 Although there is no very good agreement on the value of critical
 conductivity between the different experiments, most of the
 measurements \cite{liu} on thin films are very close to $\sigma_c
 \approx 1 (2e)^2 /h$, in a quite reasonable agreement
 with my lowest order estimate for the Coulomb universality class.
 On physical grounds, one may also 
 expect the next order term in the Eq. (11) to be again negative
 \cite{herbut1}, and that there is no second order correction in $D$ in
 the imaginary part of the self-energy \cite{herbut2}. The first
 expectation is based on the fact that disorder inhibits
 superfluidity, and therefore should effectively increase the
 exponent ($\eta$) in the power-law for the superfluid correlator
 at large distances. Also, in 1D the coupling
 $D$ measures the $2k_F$ backscattering over the random potential
 in the equivalent fermionic problem, so $D^2$ corrections should renormalize
 only the forward scattering amplitude, which in 1D is unrelated to
 conductivity. In that case the next order $O(1)$ correction in the
 Eq. (1) would be positive, presumably bringing the result closer
 to the experimental one. It is also interesting to note that for both
 universality classes the 
 critical conductivity in the hydrodynamic regime obtained  here
 turns out to be larger than the one in the coherent, $T=0$ 
 limit \cite{wallin}, \cite{herbut}. This also appears to
 be true for the SF-MI transition in a commensurate periodic
 potential \cite{damle}. Since $d=1$ represents the
 lower critical dimension for the SF-MI transition as well,
 a similar calculation to the present one could presumably
 be performed in that case, except that it would
 require a calculation of the second order in strength of the periodic
 potential. It would be interesting to compare the result obtained this way  
 with the calculation near the upper critical dimension of the ref. 12.

 Even though the Eq. (10) has been derived here with the
purpose of obtaining the universal  conductivity in $d=1+\epsilon$,
it is directly applicable to transport near the possible transition in two
edges of the incompressible quantum liquid
in a gated Hall bar \cite{arovas}. Also, by continuity, the  Eq. (10)
implies that the crossover function $F_d (x)$ is a continuously  
decreasing function of its argument for $d=1+\epsilon$, with a
maximum $\sim 1/\epsilon$ at $x=0$, and vanishing as 
$\sim \epsilon/x^{(1-\epsilon)/z}$ for large $x$. Although not
ruled out, it does seem unlikely that in $d=2$ this dependence
on $x=\hbar\omega/k_B T$ should completely disappear.
In fact, the difference in estimated
critical conductivities in completely coherent and incoherent regimes 
suggests that the situation in $d=2$ is most likely qualitatively
similar to $d=1+\epsilon$: the real part of conductivity should
continuously decrease as a function of $\hbar \omega/k_B T$,
interpolating between the two finite, $\omega=0$ and $T=0$, limits.

  Finally, the field-theoretic formulation of the renormalization group
transformation derived here has the advantage over the usual momentum-shell
calculation \cite{giamarchi}, \cite{herbut1} in that it facilitates a more
systematic higher-order calculation.  The observation that $\eta$ in $d=1$
plays a role similar to dimensionality in the classical critical
phenomena suggests a procedure analogous to  the 
standard dimensional regularization for the $d$-independent part of the
recursion relations. Adding the effect of dimensionality 
when $d>1$ as described
in ref. 14 and as done in Eqs. (11) - (13) would then yield the  
higher order corrections for the exponent $\nu$ and the critical dc
conductivity. It would be very interesting to compare the results of such
an analytical calculation with the experiments and the numerical simulations,
as it could lead to a more definite understanding of the SF-BG quantum
critical behavior.
 
  The author is grateful to NSERC of Canada and the Izaak Walton Killam
foundation for the financial support.

\pagebreak

$^*$ Permanent address.

\end{document}